\documentclass[twocolumn]{jpsj2}

\def\runauthor{
 A. \textsc{Oguri}, Y. \textsc{Nisikawa}, 
 and A. C. \textsc{Hewson}
 }

\title{
Determination of the 
phase shifts for interacting electrons 
 connected to reservoirs
}

\author{
Akira \textsc{Oguri},  
Yunori \textsc{Nisikawa}, 
and A. C. \textsc{Hewson}$^1$
}%

\inst{%
Department of Material Science, 
Osaka City University, Sumiyoshi-ku, Osaka 558-8585, Japan \\
$^1$Department of Mathematics, Imperial College, 
180 Queen's Gate, London SW7 2BZ, UK
}%

\recdate{April 29, 2005}

\abst{
We describe a formulation to deduce the phase shifts,
which determine the ground-state properties of 
interacting quantum-dot systems with the inversion symmetry, 
from the fixed-point eigenvalues 
of the numerical renormalization group (NRG). 
Our approach does not assume the specific form of the Hamiltonian 
nor the electron-hole symmetry,
and it is applicable to a wide class of quantum impurities 
connected to noninteracting leads. 
We apply the method to a triple dot which is described by  
a three-site Hubbard chain connected to two noninteracting leads, 
and calculate the dc conductance away from half-filling.
The conductance shows the typical Kondo plateaus 
of the Unitary limit $g \simeq 2e^2/h$ in some regions 
of the onsite energy $\epsilon_d$, 
at which the total number of electrons $N_{\rm el}$ in 
the three dots is odd, i.e., $N_{\rm el}\simeq 1$, $3$ and $5$. 
In contrast, the conductance shows a wide minimum 
in the regions of $\epsilon_d$ corresponding to even number of 
electrons  $N_{\rm el} \simeq 2$ and $4$. 
 We also discuss the parallel conductance of the triple dot 
 connected transversely to four leads, 
and show that it can  be deduced from the two phase shifts 
defined in the two-lead case.
}

\kword{
Quantum dot, phase shift, 
numerical renormalization group, 
Fermi liquid, Hubbard model 
}

\begin{document}
\sloppy

\maketitle

\section{Introduction}
\label{sec:introduction}

The Kondo effect in quantum dots is a subject of much current interest,
\cite{GlatzmanRaikh,NgLee}
and experimental developments in this decade make 
it possible to examine the interplay of various effects 
which have previously only been studied 
in different fields of physics.\cite{Goldharber,Cronenwett}
For instance, in quantum dots, the 
interplay of the Aharanov-Bohm, Fano, Josephson, and Kondo 
effects under equilibrium and nonequilibrium situations 
have been studied intensively.\cite{Hofstetter,Kobayashi,SC}

To study the low-temperatures properties 
of the systems showing the Kondo behavior, 
reliable theoretical approaches are required.
The Wilson numerical renormalization group (NRG) method
\cite{Wilson,KWW,KWW2} 
has been used successfully for single 
and double quantum dots.\cite{Izumida2,IzumidaK40}
In a previous work, 
we have applied the NRG method to
a finite Hubbard chain of finite size $N_C$ 
connected to noninteracting leads,
and have studied the ground-state properties at half-filling.
\cite{ao_ah}
The results obtained for $N_C=3$ and $4$  show that  
the low-lying eigenstates have one-to-one correspondence 
with the free quasi-particle excitations of a local Fermi liquid.
It enables us to determine the transport coefficients 
from the fixed-point Hamiltonian.
Furthermore, it enables us to deduce 
the characteristic parameters such as $T_K$ and Wilson ratio $R$ from 
the fixed-point eigenvalues, and the calculations have been carried out 
for the Anderson impurity model.
\cite{HewsonOguriMeyer,Hewson_JPSJ}
The purpose of this paper is to provide 
an extended formulation to deduce 
the conductance away from half-filling. 

The ground-state properties of quantum dots connected 
to two noninteracting leads are determined by 
the two phase shifts if the systems have an inversion symmetry.
Kawabata described the outline of this feature assuming that 
the low-energy properties are determined by the quasi-particles of 
a local Fermi liquid,\cite{Kawabata}
and discussed qualitatively the role of the 
Kondo resonance on the tunneling through a quantum dot.
In this paper we describe the formulation to deduce 
these two phase shifts from the fixed-point eigenvalues of NRG.
The formulation along these lines is well known for some particular cases, 
and has  been applied to the single and double quantum dots.
\cite{Hofstetter,Izumida2,IzumidaK40}
However, to our knowledge, a general description has not been given 
explicitly so far. 
Our formulation does not assume a specific form of the Hamiltonian,
and thus it is applicable to a wide class of the models for 
quantum impurities. 
We apply the method to a three-site Hubbard chain.
It is a model for a triple dot, 
and the low-energy properties of this system 
away from half-filling have not been clarified 
with reliable methods yet.
We calculate the dc conductance $g$ away from half-filling 
as a function of the onsite energy $\epsilon_d$  
which can be controlled by the gate voltage. 
The results show the typical Kondo plateaus of the 
Unitary limit $g \simeq 2e^2/h$  
when the average number of electrons $N_{\rm el}$ in the chain 
is odd, $N_{\rm el}\simeq 1$, $3$, and $5$. In contrast, 
for some finite ranges of $\epsilon_d$ corresponding 
to even number of electrons $N_{\rm el}\simeq 2$ and $4$, 
the conductance shows wide minima. 
 We also discuss the parallel conductance of the triple dot
 connected transversely to four leads.

In \S \ref{sec:MODEL},
we introduce a model for interacting electrons 
connected to reservoirs in a general form,
and describe the ground-state properties in terms of the phase shifts.   
 In \S \ref{sec:flow_vs_g},
  we describe the formulation to deduce  
 the phase shifts from the fixed-point Hamiltonian of NRG.
In \S \ref{sec:results}, we apply the formulation 
to the transport through the triple dot.
In \S \ref{sec:discussion}, 
discussion and summary are given.

\section{Model and Formulation}
\label{sec:MODEL}

\begin{figure}[bt]
\begin{center}
\leavevmode
\includegraphics[width=0.8 \linewidth]{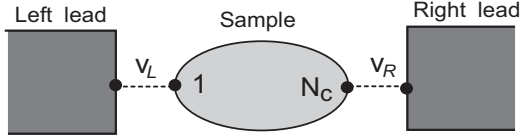}
\caption{Schematic picture of a series connection 
}
\label{fig:system}
\end{center}
\end{figure}

We start with a system consisting of 
a finite central region ($C$) 
and two leads, on the left ($L$) and the right ($R$),
as illustrated in Fig.\ \ref{fig:system}. 
The central region consists of $N_C$ quantized levels, 
and the interaction  $U_{j_4 j_3; j_2 j_1}$ is 
switched on for the electrons staying in this region.
Each of the two leads has a continuous energy spectrum. 
The complete Hamiltonian is given by 
\begin{align}
{\cal H}   & \, =\,    {\cal H}_{C}^0 \, 
+ \, {\cal H}_{C}^U 
+ \,  {\cal H}_{\rm mix} \, + \, 
{\cal H}_{\rm lead} \;,
\label{eq:H}
\\             
 {\cal H}_{C}^0  & \, = \,  
 -\sum_{ij \in C, \sigma} t_{ij}^{\phantom{0}} \, 
 d^{\dagger}_{i\sigma}d^{\phantom{\dagger}}_{j\sigma} 
\label{eq:HC^0}
 \\
 {\cal H}_{C}^U  & \, = \,  
 {1 \over 2} \sum_{\{j\} \in C}\sum_{\sigma \sigma'}    
    U_{j_4 j_3; j_2 j_1}\,  
 d^{\dagger}_{j_4 \sigma} d^{\dagger}_{j_3 \sigma'}    
 d^{\phantom{\dagger}}_{j_2 \sigma'} d^{\phantom{\dagger}}_{j_1 \sigma} 
\label{eq:HC^U}
\\
 {\cal H}_{\rm mix} \,  & \, =\,
v_{L}^{\phantom 0} 
 \sum_{\sigma}   
 \left(\,  
  d^{\dagger}_{1,\sigma} \psi^{\phantom{\dagger}}_{L \sigma}
\,+\,    
\psi^{\dagger}_{L \sigma}   d^{\phantom{\dagger}}_{1,\sigma} 
\,\right)
  \nonumber \\
& \ \ + v_{R}^{\phantom 0} 
 \sum_{\sigma}   
\left(\, 
\psi^{\dagger}_{R\sigma} d^{\phantom{\dagger}}_{N_C, \sigma} 
      \,+\,   
d^{\dagger}_{N_C, \sigma} \psi^{\phantom{\dagger}}_{R\sigma}
      \,\right)   ,
\label{eq:Hmix}
\\
 {\cal H}_{\rm lead}  & \, = \,  
\sum_{\nu=L,R} 
 \sum_{k\sigma} 
  \epsilon_{k \nu}^{\phantom{0}}\,
         c^{\dagger}_{k \nu \sigma} 
         c^{\phantom{\dagger}}_{k \nu \sigma}
\,,
\label{eq:H_lead}
\end{align}
where $d_{i\sigma}$ annihilates an electron with spin 
$\sigma$ at site $i$.
In the lead at $\nu$ ($= L,\, R$), the operator 
$c_{k \nu \sigma}^{\dagger}$ creates an electron 
with energy $\epsilon_{k\nu}$ corresponding to 
an one-particle state $\phi_{k\nu} (r)$. 
The tunneling matrix elements
 $v_L^{\phantom{\dagger}}$ 
and $v_R^{\phantom{\dagger}}$ connect 
the central region and two leads.
At the interfaces, 
a linear combination of the conduction electrons
 $\psi_{\nu \sigma}^{\phantom{\dagger}} 
= \sum_k c_{k \nu \sigma}^{\phantom{\dagger}} 
\, \phi_{k\nu} (r_{\nu}) $ mixed with the electrons at  $i=1$ or $N_C$, 
where $r_{\nu}$ denotes the position at the interface in the lead side.  
We assume that the hopping matrix elements $t_{ij}$ 
and $v_{\nu}^{\phantom{\dagger}}$ are real,
and the interaction has the time-reversal symmetry:  
$U_{4 3; 2 1}$ is real 
and $U_{4 3; 2 1}=U_{3 4; 1 2}=U_{1 2; 3 4 }=U_{4 2; 3 1}=U_{1 3; 2 4}$. 
We will be using units $\hbar=1$ unless otherwise noted.

We use the Green's function for 
the interacting region ($i j \in C$) defined by
\begin{equation} 
G_{ij}(\text{i}\omega_n) 
\, =\,  
-    \int_0^{\beta} \! \text{d}\tau \,
   \left \langle  T_{\tau} \,  
   d^{\phantom{\dagger}}_{i \sigma} (\tau) 
   \, d^{\dagger}_{j \sigma} (0) 
     \right \rangle  \, \text{e}^{\text{i} \omega_n \tau} \;,
\label{eq:G_Matsubara}
\end{equation} 
where $\beta= 1/T$,  
$d_{j \sigma}(\tau) = 
\text{e}^{\tau  {\cal H}} d_{j \sigma} \text{e}^{- \tau  {\cal H}}$, 
and $\langle {\cal O} \rangle =
\mbox{Tr} \left[ \, \mathrm{e}^{-\beta  {\cal H} }\, {\cal O}
\,\right]/\mbox{Tr} \, \mathrm{e}^{-\beta  {\cal H} }$. 
The corresponding retarded and advanced functions are given by 
 $G_{ij}^{\pm}(\omega) =  G_{ij}(\omega \pm \text{i}\, \delta)$ 
via the analytic continuation. 
The Dyson equation for the Green's function 
  $\mbox{\boldmath $G$}(z) = \{G_{ij}(z)\}$  
can be rewritten in an $N_C \times N_C$ matrix form 
\begin{equation}
\left\{\mbox{\boldmath $G$}(z)\right\}^{-1}  
  \, = \,   
z \, \mbox{\boldmath $1$} 
-  \mbox{\boldmath $H$}_C^0  
- \mbox{\boldmath $V$}_{\rm mix}(z)   
- \mbox{\boldmath $\Sigma$}(z)   \;,
\label{eq:G_matrix}
\end{equation}
where $\mbox{\boldmath $H$}_C^0 = \{-t_{ij}^{\phantom{0}} \}$, 
\begin{align}
\mbox{\boldmath $V$}_{\rm mix}(z) &\,= \, 
 \left [ \, 
\begin{matrix}
 v_L^{2} \mbox{\sl g}_L^{\phantom{\dagger}}(z) & 0 & & \mbox{\Large $0$} \cr
  0   &    0     & \ \ddots   &      \cr
       &  \ddots \ & \ddots   &  0  \cr 
 \mbox{\Large $0$} & & 0 & v_R^{2} \mbox{\sl g}_R^{\phantom{\dagger}}(z) \cr
\end{matrix}
 \, \right ]  
\label{eq:V_mix} 
\;,
\end{align}
and $\mbox{\boldmath $\Sigma$}(z) = \{\Sigma_{ij}(z)\}$  
is the self-energy due to the inter-electron interaction ${\cal H}_C^U$.  
In the matrix $\mbox{\boldmath $V$}_{\rm mix}(z)$,  
the two non-zero elements are determined by 
the Green's function at interface of the isolated lead
 $\mbox{\sl g}_{\nu}^{+}(\omega) \equiv \sum_k
{
\left| \phi_{\nu \sigma}^{\phantom{\dagger}} (r_{\nu}) \right|^2
/ 
(\omega - \epsilon_{k\nu}+ i0^+) 
}$. 
We assume, for simplicity, 
that the density of states $\rho_{\nu}$ is a constant 
for small $\omega$. Then, 
the energy scale of the level-broadening 
becomes $\Gamma_{\nu}= \pi\rho_{\nu} v_{\nu}^2$, and 
$\mbox{\sl g}_{\nu}^{+}(\omega) = -\text{i}\pi \rho_{\nu}$.

\begin{figure}[bt]
\begin{center}
\leavevmode
\includegraphics[width=0.7 \linewidth]{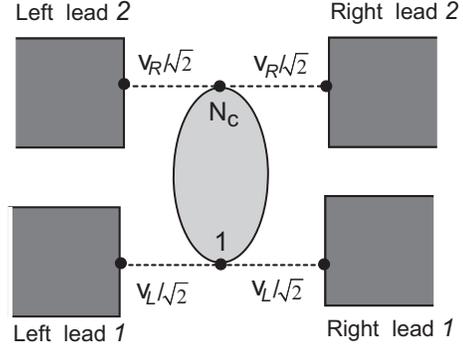}
\caption{Schematic picture of a parallel connection.
}
\label{fig:parallel}
\end{center}
\end{figure}

\subsection{Ground state properties}

If the self-energy shows a 
property $\mbox{Im}\, \mbox{\boldmath $\Sigma$}^{+} (0) =0$  at $T=0$, 
then the damping of the single-particle excitations 
at the Fermi level vanishes. 
In this case, the renormalization hopping matrix elements 
$\mbox{\boldmath $H$}_C^{\rm eff} =\{ -\widetilde{t}_{ij}\, \}$ 
defined by
\begin{equation}
\mbox{\boldmath $H$}_C^{\rm eff}  
\, \equiv\, 
   \mbox{\boldmath $H$}_C^0  
            + \mbox{Re}\, \mbox{\boldmath $\Sigma$}^+(0)
\label{eq:K} 
\end{equation}
play a central role on the ground-state properties.
The value of the Green's function 
at the Fermi level is given by 
$\left\{\mbox{\boldmath $G$}^+(0)\right\}^{-1}  = 
\mbox{\boldmath $K$}(0)  -  \mbox{\boldmath $V$}_{\rm mix}^+(0)$ with  
 \begin{equation}
\mbox{\boldmath $K$}(\omega) \,\equiv\,  
\omega \,\mbox{\boldmath $1$} 
-    
\mbox{\boldmath $H$}_C^{\rm eff}  
\;, 
\label{eq:K_omega} 
\end{equation}
and the scattering coefficients at $T=0$ are 
determined by the determinant  
\begin{align}
\det \left\{\mbox{\boldmath $G$}^+(0)\right\}^{-1} 
 & \,=\,         \left[ \, 
        -  \Gamma_L \,\Gamma_R \, \det \mbox{\boldmath $K$}_{11}^{N_CN_C}(0) 
        + \det \mbox{\boldmath $K$}(0) 
       \, \right] \, 
      \nonumber \\
&
\!\!\!\!\!\!
+ \, \text{i}\, 
        \Bigl[ \,\Gamma_L \, \det \mbox{\boldmath $K$}_{11}(0) + 
               \Gamma_R \, \det \mbox{\boldmath $K$}_{N_CN_C}(0)
        \, \Bigr] 
\;.
\label{eq:determinant_Ginv}
\end{align}
Here $\mbox{\boldmath $K$}_{ij}(0)$ is 
an $(N_C-1) \times (N_C-1)$  derived 
from $\mbox{\boldmath $K$}(0)$ by 
deleting the $i$-th row and  $j$-th column.
Similarly, $\mbox{\boldmath $K$}_{11}^{N_CN_C}(0)$ is 
an $(N_C-2) \times (N_C-2)$ matrix obtained from $\mbox{\boldmath $K$}(0)$ 
by deleting the first and $N_C$-th rows, 
and the first and $N_C$-th columns. 
The dc conductance at $T=0$ can be expressed, using 
an inter-boundary Green's function,\cite{ao8,ao9,ao10}
as 
\begin{equation}
g_{\rm series}^{\phantom{0}}\, = \, \frac{2 e^2}{h}  
    4\Gamma_R \Gamma_L \left| G_{N_C 1}^{+}(0)\right|^2  
         \;. 
\label{eq:cond} 
\end{equation}
Similarly, at $T=0$ the charge displacement $N_{\rm el}$ 
caused by the hybridizations $\Gamma_L$ and $\Gamma_R$ 
can be expressed, using 
the Friedel sum rule,\cite{LangerAmbegaokar,tanaka1}  as 
\begin{align}
N_{\rm el} &\,=\,
{1 \over \pi \text{i} }\,   
\log \left[ \det \mbox{\boldmath $S$} \right]
\;,
\label{eq:Friedel}
\\
\mbox{\boldmath $S$}
&\,= 
  \left [ \,
 \begin{matrix}  
 1 -2\text{i}\,\Gamma_L G_{1 1}^{+}(0) 
 & \ \ -2\text{i}\,\Gamma_L G_{1 N_C}^{+}(0)    \cr
 -2\text{i}\,\Gamma_R G_{N_C 1 }^{+}(0)  
 &  \rule{0cm}{0.5cm} \ 1 -2\text{i}\,\Gamma_R G_{N_C N_C }^{+}(0) \cr  
 \end{matrix}
 \, \right ]  
\;.
\label{eq:S}
\end{align}
In the case of the constant density of states we are assuming, 
the charge displacement coincides with the total number of electrons 
in the central region, i.e.,     
$N_{\rm el} =  
\sum_{i=1}^{N_C} \sum_{\sigma}
 \langle n_{d,i\sigma}^{\phantom{0}}\rangle$,  
where  $n_{d,i\sigma}^{\phantom{0}}= 
d_{i\sigma}^{\dagger} d_{i\sigma}^{\phantom{\dagger}}$.
Note that the wavefunction renormalization due to the matrix
 $\mbox{\boldmath $1$}-\partial 
\mbox{\boldmath $\Sigma$}/(\partial \omega)$ 
does not contribute to the conductance and charge displacement 
at $T=0$, although it has an information about $T_K$ 
and can be related to the asymptotic behavior of the low-lying 
excited states near the fixed-point of NRG.
\cite{HewsonOguriMeyer,Hewson_JPSJ,Hewson_jphys}

Using the above Green's functions obtained for the series connection, 
one can calculate the conductance for the parallel connection, 
for which the tunneling matrix elements are given by 
$v_L/\sqrt{2}$ and $v_R/\sqrt{2}$ for 
the leads labeled by \lq\lq 1" and  \lq\lq 2", 
respectively as shown in Fig.\ \ref{fig:parallel}.
Owing to this special symmetry between the left and right leads,
the interacting site at $i=1$ ($i=N_C$) is coupled only to 
an even combination of the states from the left 
and right leads with the label \lq\lq 1" (\lq\lq 2").
Thus, the Green's functions in the central region 
$\mbox{\boldmath $G$}(z)$ coincide with that for the series connection,
so that a ground-state average such as $N_{\rm el}$ becomes 
the same with 
that for the series connection.
The parallel conductance, however, 
 is different because two conducting channels contribute to 
the total current flowing in the horizontal direction,
and it can be expressed in the form
\begin{align}
g_{\rm parallel}^{\phantom{0}} =  
{2 e^2 \over h} 
& \, 2 
\Bigl[\,
\Gamma_L^2 \, \left|G_{11}^+(0) \right|^2
+ 
\Gamma_R^2 \, \left|G_{N_C N_C}^+(0) \right|^2
\nonumber
\\
& \quad
+ \,  2 
\Gamma_L 
\Gamma_R \, \left|G_{N_C 1}^+(0) \right|^2
\,\Bigr] \;.
\label{eq:g_parallel}
\end{align}
This expression is for the current flowing 
from left to right (or right to left),  
and has been derived from the Kubo formula 
using a multi-channel formulation for the dc conductance. 
\cite{tanaka1}

\subsection{Inversion symmetric case}

We now consider the case the system has an inversion symmetry;  
 $\Gamma_L = \Gamma_R$ ($\equiv \Gamma$), 
 $\mbox{\sl g}_{L} = \mbox{\sl g}_{R}$  ($\equiv \mbox{\sl g}$), 
 and $t_{i,j} =t_{N_C-i+1,N_C-j+1}$. 
Due to this symmetry,
the eigenstates of the Hamiltonian ${\cal H}$
can be classified according to the parity, 
and the matrix Dyson equation given in eq.\ (\ref{eq:G_matrix}) can be 
separated into the two subspaces 
corresponding to the even and odd orbitals;
\begin{align}
a_{j,\sigma} &\,= {
d_{j,\sigma} \,+\, d_{N_C-j+1,\sigma} 
\over \sqrt{2}
 } \;,
 \label{eq:orbitals_a}
 \\
b_{j,\sigma} &\,= {
d_{j,\sigma} \,-\, d_{N_C-j+1,\sigma} 
\over \sqrt{2} } \;,
 \label{eq:orbitals_b}
\end{align}
where $j=1,2, \ldots , N_C/2$ for even $N_C$, and 
  $j=1,2, \ldots , (N_C-1)/2$  for odd $N_C$.
Note that there is one extra unpaired orbitals 
$a_{(N_C+1)/2,\sigma} \equiv d_{(N_C+1)/2,\sigma}$ for odd $N_C$.
Thus the Green's functions for 
these orbitals $a_{j,\sigma}$ and $b_{j,\sigma}$ 
can be written separately in the forms 
\begin{align}
\left\{
\mbox{\boldmath $G$}^+_{\rm even}(0) \right\}^{-1}
& \, = \, 
\mbox{\boldmath $K$}^{\rm even}(0)  
- \mbox{\boldmath $V$}_{\rm mix}^{\rm even}(0)   
\;,
 \label{eq:Dyson_even}
\\
\left\{
\mbox{\boldmath $G$}^+_{\rm odd}(0) \right\}^{-1}
& \, = \, 
\mbox{\boldmath $K$}^{\rm odd}(0)  
- \mbox{\boldmath $V$}_{\rm mix}^{\rm odd}(0)   
\;,
 \label{eq:Dyson_odd}
\end{align}
at $T=0$ and $\omega=0$. 
The matrices 
$\mbox{\boldmath $K$}^{\rm even}(0)$ and 
$\mbox{\boldmath $K$}^{\rm odd}(0)$ are derived from 
$\mbox{\boldmath $K$}(0)$ in eq.\ (\ref{eq:K_omega}) 
via the transformation 
 eqs.\ (\ref{eq:orbitals_a}) and (\ref{eq:orbitals_b}).
Since only the even and odd orbitals with the label $j=1$, 
i.e., $a_{1,\sigma}$ and $b_{1,\sigma}$, 
have a finite tunneling-matrix element to the lead with the same parity, 
the mixing term defined in eq.\ (\ref{eq:V_mix}) is transformed as
\begin{align}
\mbox{\boldmath $V$}_{\rm mix}^{\gamma}(\omega) &\,  =  \,
 \left [ \, 
\begin{matrix}
 v^{2} \mbox{\sl g}^{+}(\omega) & \mbox{\Large $0$} \cr
\mbox{\Large $0$}  & \mbox{\Large $0$} \cr
\end{matrix}
 \, \right ]   
 \;,
 \end{align}
for $\gamma=$ \lq\lq even" and \lq\lq odd". 
Note that the size of 
 $\mbox{\boldmath $V$}_{\rm mix}^{\rm even}$ and 
 that of  $\mbox{\boldmath $V$}_{\rm mix}^{\rm odd}$ are 
 identical for even $N_C$, whereas 
the size of these two matrices are 
 different for odd $N_C$ because the one unpaired orbitals exists.
The determinant given in eq.\ (\ref{eq:determinant_Ginv}) can be 
factorized, as 
\begin{align}
 \det \left\{\mbox{\boldmath $G$}^+(0) \right\}^{-1} 
 &\, = \,
 \det \left\{\mbox{\boldmath $G$}^+_{\rm even}(0) \right\}^{-1}  \, 
 \det \left\{\mbox{\boldmath $G$}^+_{\rm odd}(0) \right\}^{-1} 
\nonumber
\\
& \,=\, 
\prod_{\gamma}       
\Bigl[ \, 
  \det \mbox{\boldmath $K$}^{\gamma}(0) \,+\,
 \text{i}\,\Gamma\, \det \mbox{\boldmath $K$}^{\gamma}_{11}(0) 
 \,\Bigr]
\;, 
\label{eq:detG_parity}
\end{align}
where the product runs over the two partial 
waves $\gamma=$ \lq\lq even" and \lq\lq odd". 
The matrices $\mbox{\boldmath $K$}^{\rm even}_{11}(0)$ and 
$\mbox{\boldmath $K$}^{\rm odd}_{11}(0)$  are 
 derived respectively from 
$\mbox{\boldmath $K$}^{\rm even}(0)$ and 
$\mbox{\boldmath $K$}^{\rm odd}(0)$ 
by deleting the first row and column 
corresponding to the orbitals  $a_{1,\sigma}$ and $b_{1,\sigma}$, 
which are connected directly to noninteracting leads.
The scattering coefficients  are determined by the ratios,
$\kappa_{\rm even}$ and $\kappa_{\rm odd}$, defined by  
\begin{align}
\kappa_{\gamma} 
& \, \equiv \, 
{
\det \mbox{\boldmath $K$}^{\gamma}(0) 
\over
\Gamma \det \mbox{\boldmath $K$}^{\gamma}_{11}(0)
} 
\;.
\label{eq:kappa2_def}
\end{align}
The value of $\kappa_{\rm even}$ and  $\kappa_{\rm odd}$ 
 can be deduced from the fixed-point eigenvalues of NRG,
as described in eq.\ (\ref{eq:kappa2}) in \S \ref{sec:flow_vs_g}.

The phase shifts for the even and odd partial waves, 
$\delta_{\rm even}$ and $\delta_{\rm odd}$, 
can be defined using the retarded Green's 
functions for $a_{1,\sigma}$ and $b_{1,\sigma}$ at the Fermi level, 
which can be expressed, using eq.\ (\ref{eq:detG_parity}), as 
\begin{align}
 \langle\langle a_{1,\sigma}^{\phantom{\dagger}}; 
 a_{1,\sigma}^{\dagger} \rangle\rangle_{\omega=0}
&
\, = \, \frac{1}{\Gamma} \, \frac{1}{\kappa_{\rm even} + \text{i} } 
\; \,= \,    
\frac{1}{\Gamma} \, 
\frac{\text{e}^{\text{i} \delta_{\rm even}}}{\sqrt{\kappa_{\rm even}^2 + 1}}
\;,
\label{eq:G_even}
\\
 \langle\langle b_{1,\sigma}^{\phantom{\dagger}}; 
 b_{1,\sigma}^{\dagger} \rangle\rangle_{\omega=0}
&
\, = \, \frac{1}{\Gamma} \, \frac{1}{\kappa_{\rm odd} + \text{i}}
\; \,=\,
\frac{1}{\Gamma} \, 
\frac{\text{e}^{\text{i} \delta_{\rm odd}}}{\sqrt{\kappa_{\rm odd}^2 + 1}} 
\;.
\label{eq:G_odd}
\end{align}
The Green's functions at the interfaces 
$G_{N_C, 1}^+(0)$ and $G_{11}^+(0)$ are given by 
the linear combinations
\begin{align}
G_{N_C, 1}^+(0)  
   &\,=\, \frac{1}{2 \Gamma} 
   \left[\, 
   \frac{1}{\kappa_{\rm even} + \text{i}}
   \,-\, \frac{1}{\kappa_{\rm odd} + \text{i}}
   \,\right] \;,
\label{eq:G_N1_sym}
\\
G_{11}^+(0) 
    &\,=\, \frac{1}{2 \Gamma} 
    \left[\, 
    \frac{1}{\kappa_{\rm even} + {i}}
    \,+\, \frac{1}{\kappa_{\rm odd} + \text{i}}
    \,\right] \;.
\label{eq:G_11_sym}
\end{align}
Thus, the conductance defined in eq.\ (\ref{eq:cond}) 
can be  written in the form 
\begin{align}
g_{\rm series}^{\phantom{0}} 
&\,= \, 
{2 e^2 \over h} \, \sin^2 
\Bigl( \delta_{\rm even} - \,\delta_{\rm odd} \Bigr)
\;.
 \label{eq:g_sym}
\end{align}
Similarly, 
the Friedel sum rule eq.\ (\ref{eq:Friedel}) can be rewritten in the form 
\begin{align}
 \sum_{i=1}^{N_C} \sum_{\sigma} \langle 
 n_{d,i\sigma}^{\phantom{0}} \rangle 
&\,= \, \frac{2}{\pi}\, 
\Bigl( \delta_{\rm even} + \,\delta_{\rm odd} \Bigr)
\;.
\label{eq:Friedel_inv}
\end{align}
Furthermore, 
the parallel conductance for the current flowing in 
the horizontal direction in 
the geometry shown in Fig.\ \ref{fig:parallel} 
can be expressed using these two phase shifts, i.e., 
the inversion symmetry  simplifies eq.\ (\ref{eq:g_parallel}) as
\begin{align}
g_{\rm parallel}^{\phantom{0}} 
&\,=\, 
{2 e^2 \over h} \, 
\Bigl( 
\sin^2 
 \delta_{\rm even} + \sin^2  \,\delta_{\rm odd} \Bigr)
\;.
\label{eq:cond_para2}
\end{align}
Thus the parallel conductance is a simple sum of the contributions of  
the even and odd channels.\cite{Izumida2,IzumidaK40}
Alternatively, 
it can be expressed in the form, $g_{\rm parallel}^{\phantom{0}} =
(2 e^2 / h) \, 2 \Gamma 
\left\{ -\mbox{Im}\, G_{11}^+(0) \right\}$.

\section{Fixed-point Hamiltonian and phase shifts}
\label{sec:flow_vs_g}

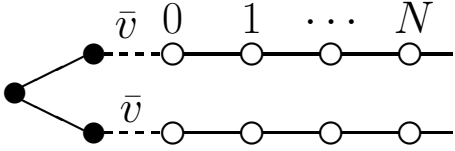
\begin{figure}[bt]
\begin{center}

\setlength{\unitlength}{0.7mm}
\begin{picture}(130,40)(-12,0)
\thicklines


\put(30,12){\makebox(0,0)[bl]{\Large$\bar{v}$}}
\put(29,28){\makebox(0,0)[bl]{\Large$\bar{v}$}}

\put(38,29){\makebox(0,0)[bl]{\Large$0$}}
\put(53,29){\makebox(0,0)[bl]{\Large$1$}}
\put(65,29){\makebox(0,0)[bl]{\Large$\cdots$}}
\put(82,29){\makebox(0,0)[bl]{\Large$N$}}



\put(9,17.5){\line(2,1){14}}
\put(9,17.5){\line(2,-1){14}}
\put(10,17.5){\circle*{4}}


\multiput(27,10)(3,0){4}{\line(1,0){1.5}}
\put(42,10){\line(1,0){11}}
\put(57,10){\line(1,0){11}}
\put(72,10){\line(1,0){11}}
\put(87,10){\line(1,0){6}}

\multiput(27,25)(3,0){4}{\line(1,0){1.5}}
\put(42,25){\line(1,0){11}}
\put(57,25){\line(1,0){11}}
\put(72,25){\line(1,0){11}}
\put(87,25){\line(1,0){6}}

\put(25,10){\circle*{4}} 
\put(40,10){\circle{4}} 
\put(55,10){\circle{4}} 
\put(70,10){\circle{4}} 
\put(85,10){\circle{4}} 

\put(25,25){\circle*{4}} 
\put(40,25){\circle{4}} 
\put(55,25){\circle{4}} 
\put(70,25){\circle{4}} 
\put(85,25){\circle{4}} 
\end{picture}

\caption{Schematic pictures of discretized Hamiltonian of NRG.}
\label{fig:model_NRG}
\end{center}

\end{figure}

In the NRG approach the conduction band is transformed into 
 a linear chain as shown in Fig.\ \ref{fig:model_NRG} 
 after carrying out a standard procedure of logarithmic discretization.
\cite{Wilson,KWW,KWW2}
Then, a sequence of the Hamiltonian $H_N$ is introduced, as
\begin{align}
H_N & \,=\,  \Lambda^{(N-1)/2}  
\left( \,
      {\cal H}_{C}^0  + {\cal H}_{C}^U 
 +  H_{\rm mix}^{\phantom{0}}  +   H_{\rm lead}^{(N)}
 \,\right) ,
\label{eq:H_N} 
\\
H_{\rm mix}^{\phantom{0}} & \, = \, \bar{v} \, 
       \sum_{\sigma}
\left(\,
f^{\dagger}_{0,L\sigma} d^{\phantom{\dagger}}_{ 1,\sigma}
\,+\, 
d^{\dagger}_{ 1,\sigma}  f^{\phantom{\dagger}}_{0,L\sigma} 
     \right)
     \nonumber \\
 & \quad  +   
        \bar{v}\, 
       \sum_{\sigma} 
       \left(\,
f^{\dagger}_{0,R \sigma} d^{\phantom{\dagger}}_{N_C, \sigma} 
   \,+\, 
d^{\dagger}_{N_C, \sigma} f^{\phantom{\dagger}}_{0,R \sigma} 
         \,  \right)  \;,
\label{eq:H_mix_NRG}
\\
H_{\rm lead}^{(N)} &\,=\,
D\,{1+1/\Lambda \over 2} \,
\sum_{\nu=L,R}
\sum_{\sigma}
\sum_{n=0}^{N-1} 
\, \xi_n\, \Lambda^{-n/2}
\nonumber \\
& \qquad \quad \times  
\left(\,
  f^{\dagger}_{n+1,\nu\sigma}\,f^{\phantom{\dagger}}_{n,\nu\sigma}
  +  
 f^{\dagger}_{n,\nu\sigma}\, f^{\phantom{\dagger}}_{n+1,\nu\sigma}
 \,\right) \;,
\label{eq:H_lead_NRG}
\end{align}
where $D$ is the half-width of the conduction band.
The hopping matrix elements   
  $\bar{v}$ and 
 $\xi_n$  are defined by
\begin{align}
  \bar{v}
&\,=\, \sqrt{ \frac{2D\,\Gamma A_{\Lambda}}{\pi} }
\;,
\qquad
A_{\Lambda}  \,=\,  \frac{1}{2}\, 
 {1+1/\Lambda \over 1-1/\Lambda }
\,\log \Lambda
\;, 
\label{eq:A_lambda}
\\
\xi_n &\,=\,    
{ 1-1/\Lambda^{n+1}  
\over  \sqrt{1-1/\Lambda^{2n+1}}  \sqrt{1-1/\Lambda^{2n+3}} 
} 
\;.
\label{eq:xi_n}
\end{align}
The factor $A_{\Lambda}$ is required for comparing precisely 
the discretized Hamiltonian  $H_N$ with the original continuous model 
defined in eq.\ (\ref{eq:H}).\cite{KWW,SakaiShimizuKasuya}
In the discretized Hamiltonian $H_N$,  the matrix elements 
$t$ and $\bar{v}$ are multiplied by $\Lambda^{(N-1)/2}$,
so that the original Hamiltonian ${\cal H}$ are recovered  
from $\Lambda^{-(N-1)/2} H_N$ 
in the limit of $N \to \infty$ and $\Lambda \to 1$.

In a wide class of impurity models described by eq.\ (\ref{eq:H}),
the low-lying energy eigenvalues of $H_N$ converge for large $N$ 
to the spectrum which has one-to-one correspondence to 
the quasi-particles of a local Fermi liquid.\cite{KWW,KWW2,ao_ah}
It enables us to deduce the matrix elements of 
$\mbox{\boldmath $H$}_C^{\rm eff}$ from the NRG spectrum.
At the fixed point, 
the low-energy spectrum of the many-body Hamiltonian 
$H_N$ can be reproduced by the one-particle Hamiltonian consisting 
of $\mbox{\boldmath $H$}_{C}^{\rm eff}$ 
and the two finite leads;  
\begin{equation}
H_{\rm qp}^{(N)}  \, = \, 
\Lambda^{(N-1)/2}  
\left( \,
H_{C}^{\rm eff} + H_{\rm mix} + H_{\rm lead}^{(N)} 
\,\right) 
\;.
\label{eq:H_qp}
\end{equation}
Here  $H_{C}^{\rm eff} \equiv  - \sum_{ij=1}^{N_C} 
\widetilde{t}_{ij} \, 
d^{\dagger}_{i \sigma} d^{\phantom{\dagger}}_{j \sigma}$, 
and $\widetilde{t}_{ij}$ is 
the renormalized matrix element defined in eq.\ (\ref{eq:K}).
The Hamiltonian $H_{\rm qp}^{(N)}$  describes 
the free quasi-particles in the system 
consisting of $N_C + 2(N+1)$ sites,  
and the corresponding Green's function 
for the interacting sites can be written 
as an $N_C \times N_C$ matrix, 
\begin{align}
& \left\{\mbox{\boldmath $G$}_{\rm qp}(\omega)\right\}^{-1} 
\, \equiv \,
\nonumber \\
& 
\Lambda^{(N-1)/2} \,  
 \Bigl[ \,
 \omega  \Lambda^{-(N-1)/2} \,\mbox{\boldmath $1$} -
 \mbox{\boldmath $H$}_C^{\rm eff}  
 -  \Lambda^{(N-1)/2} \, \mbox{\boldmath $V$}_{\rm mix}^+(\omega) 
\, \Bigr ]  .
\label{eq:G_qp}
\end{align}
Here we have not included the renormalization 
factor $\partial \mbox{\boldmath $\Sigma$}/\partial \omega$ 
in eq.\ (\ref{eq:G_qp}),
because at $T=0$ it does not affect the dc conductance 
and charge displacement $N_{\rm el}$.
The eigenvalue $\varepsilon^*$ of $H_{\rm qp}^{(N)}$ satisfies the condition  
$\det \left\{\mbox{\boldmath $G$}_{\rm qp}(\varepsilon^*)\right\}^{-1} = 0$. 
Owing to the inversion symmetry, 
this equation can be factorized, as eq.\ (\ref{eq:detG_parity})  
\begin{align}
&\prod_{\gamma} \left[\,
 \det \mbox{\boldmath $K$}^{\gamma}(\omega_N) \,-\,
\bar{v}^2 
 \Lambda^{(N-1)/2}  
  \mbox{\sl g}_N^{\phantom{\dagger}}(\varepsilon^*) 
\, \det \mbox{\boldmath $K$}^{\gamma}_{11}(\omega_N) 
\,\right]
\nonumber
\\
& \,= \,0 \;,
\label{eq:NRG_eigenvalue}
\end{align}
where 
the product runs over the two partial waves,  
$\gamma=$ \lq\lq ${\rm even}$" and \lq\lq ${\rm odd}$". 
In eq.\ (\ref{eq:NRG_eigenvalue}), 
the argument for the determinants is 
$\omega_N  \equiv \varepsilon^*\,\Lambda^{-(N-1)/2}$,
 which becomes zero in the limit of  $N\to \infty$. 
The Green's function  $\mbox{\sl g}_N (\omega)$ is defined 
in the site $n=0$ at the interface 
of an isolated lead consisting of $N+1$ sites,
and it can  be expressed in the form 
 $\mbox{\sl g}_N (\omega) = 
\sum_{m=0}^{N} |\varphi_m(0)|^2/(\omega- \epsilon_m)$,
where $\epsilon_m$ and $\varphi_m(n)$ 
are the eigenvalue and eigenfunction for the isolated lead.
Furthermore, 
using eq.\ (\ref{eq:NRG_eigenvalue}),
the eigenstates of $H_{\rm qp}^{(N)}$ can be classified 
according to the parity. 
Thus, the two parameters $\kappa_{\rm even}$ and  $\kappa_{\rm odd}$,
which are defined in eq.\ (\ref{eq:kappa2_def}),
can be deduced from the low-lying eigenvalues 
for the quasi-particles of the even and odd parities, 
 $\varepsilon^*_{\rm even}$ and  $\varepsilon^*_{\rm odd}$,  as
\begin{align}
&
\kappa_{\gamma} \,\equiv \,
{
\det \mbox{\boldmath $K$}^{\gamma}(0) 
\over
\Gamma \det \mbox{\boldmath $K$}^{\gamma}_{11}(0)
} \,= \,
\frac{\bar{v}^2}{\Gamma D} 
 \lim_{N\to \infty}
 D \, 
\Lambda^{(N-1)/2}  
 \mbox{\sl g}_N^{\phantom{\dagger}}(\varepsilon^*_{\gamma}) 
\; 
\label{eq:kappa2}
\end{align}
for $\gamma=$ \lq\lq ${\rm even}$" and \lq\lq ${\rm odd}$". 
Note that
 $\bar{v}^2/(\Gamma D)  =  2 A_{\Lambda} / \pi$ 
from the definition eq.\ (\ref{eq:A_lambda}).
With these eigenstates,
the quasi-particle Hamiltonian is written in a diagonal form 
\begin{equation}
H_{\rm qp}^{(N)}  
= 
\sum_{\sigma}\sum_{l \gamma}
\left( 
\varepsilon_{p,l \gamma}^* \,
\alpha_{l\gamma\sigma}^{\dagger} \alpha_{l\gamma\sigma}^{\phantom{\dagger}}
-\, 
\varepsilon_{h,l \gamma}^* \,
\beta_{l\gamma\sigma}^{\dagger} \beta_{l\gamma\sigma}^{\phantom{\dagger}}
\right)  \, + E_g \,,
\label{eq:Hqp_even}
\end{equation}
where $E_g$ is the ground-state energy, and
$\varepsilon_{p,l \gamma}^*$ ($\varepsilon_{h,l \gamma}^*$) is 
the $l$-th excitation energy of 
a single-particle (single-hole) state 
for parity $\gamma$.
Thus, 
the parameters $\kappa_{\rm even}$ and $\kappa_{\rm odd}$ 
can be calculated substituting the NRG result of 
a low-energy eigenvalue, 
$\varepsilon_{p,l\gamma}^*$ or $-\varepsilon_{h,l\gamma}^*$, 
into the right-hand side of eq.\ (\ref{eq:kappa2}).
Then, 
the conductance, local charge, and  
phase shifts $\delta_{\rm even}$ and $\delta_{\rm odd}$, 
can be calculated using the equations given in \S \ref{sec:MODEL}.

\begin{figure}[bt]

\begin{center}

\setlength{\unitlength}{0.7mm}

\begin{picture}(130,20)(-5,0)
\thicklines

\put(11,4){\line(1,0){25}}
\put(11,16){\line(1,0){25}}
\put(36,4){\line(0,1){12}}

\put(84,4){\line(1,0){25}}
\put(84,16){\line(1,0){25}}
\put(84,4){\line(0,1){12}}

\multiput(37,10)(2,0){6}{\line(1,0){1}}
\multiput(73,10)(2,0){6}{\line(1,0){1}}

\put(48,10){\circle*{5}} 
\put(60,10){\circle*{5}} 
\put(72,10){\circle*{5}} 

\put(50,10){\line(1,0){10}}
\put(60,10){\line(1,0){10}}

\put(54,15){\makebox(0,0)[bl]{\Large $t$}}
\put(64,15){\makebox(0,0)[bl]{\Large $t$}}
\put(41,15){\makebox(0,0)[bl]
{\Large $v$}}
\put(76,15){\makebox(0,0)[bl]
{\Large $v$}}


\end{picture}
\caption{Schematic picture of a series connection for $N_C=3$.}
\label{fig:series}
\end{center}
\end{figure}
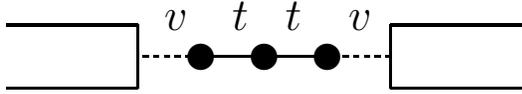

Alternatively, one can calculate the conductance 
directly from the current-current correlation function 
without using the fixed-point Hamiltonian of the local Fermi liquid.
\cite{Izumida2,IzumidaK40}
It is more general, 
and is applicable also at finite temperatures.
For the systems consisting of a number of interacting sites $N_C$,
 however, the number of eigenstates needed   
for carrying out the NRG iterations becomes large,
and the calculations of the expectation values 
require a rather long computer-time. Therefore,
for studying the ground-state properties, 
the formulation described above is much easier.
Moreover, our formulation requires only the eigenvalues to deduce  
the phase shifts via eq.\ (\ref{eq:kappa2}).
This is more efficient than deducing them from the
expectation values of the local charge $\langle n_{d,i\sigma}\rangle$,
which would have to be determined by an additional calculation.

\section{Phase shifts for triple-dot systems}
\label{sec:results}

We now apply the formulation described in the above 
to a finite Hubbard chain coupled to two noninteracting leads.
Specifically, we concentrate on a triple dot 
as illustrated in Fig.\ \ref{fig:series}.
The explicit form of the interaction Hamiltonian is 
$
{\cal H}_C^U = U \sum_{i=1}^{N_C} 
   n_{d,i \uparrow}^{\phantom{0}}\, n_{d,i \downarrow}^{\phantom{0}}
$ with $N_C = 3$.
We assume that 
the bare hopping matrix elements $t_{ij}$, 
which are defined in eq.\ (\ref{eq:HC^0}) for ${\cal H}_C^0$,
are described by the nearest-neighbor hopping $t$, 
and assume that all other off-diagonal elements to be zero.
The onsite energy of the interacting cites 
 is given by $-t_{ii}=\epsilon_d$  for $1\leq i \leq N_C$, 
and the origin of the energy is chosen to be $\mu=0$.

\begin{figure}[tb]
\begin{center}
\leavevmode
\includegraphics[ width=0.9\linewidth]{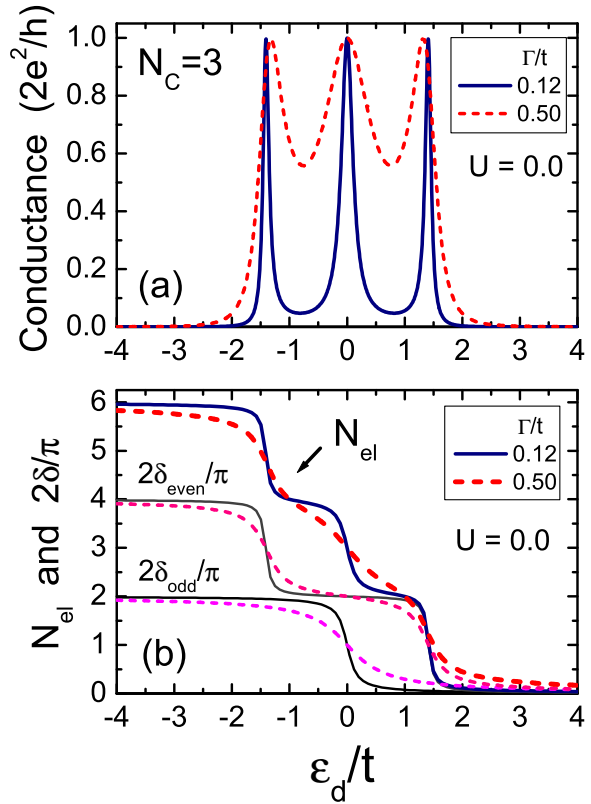}
\caption{
The conductance $g_{\rm series}$, and 
total charge in the triple-dot $N_{\rm el}$
in the noninteracting case $U=0$ as functions of $\epsilon_d/t$ 
for (solid line) $\Gamma/t = 0.12$ and (dashed line) $\Gamma/t=0.5$.  
The phase shifts  
 $2\delta_{\rm even}/\pi$ and 
 $2\delta_{\rm odd}/\pi$ are also shown 
in (b).
}
\label{fig:cond_u0}
\end{center}
\end{figure}

\begin{figure}[tb]
\begin{center}
\leavevmode
\includegraphics[ width=0.9\linewidth]{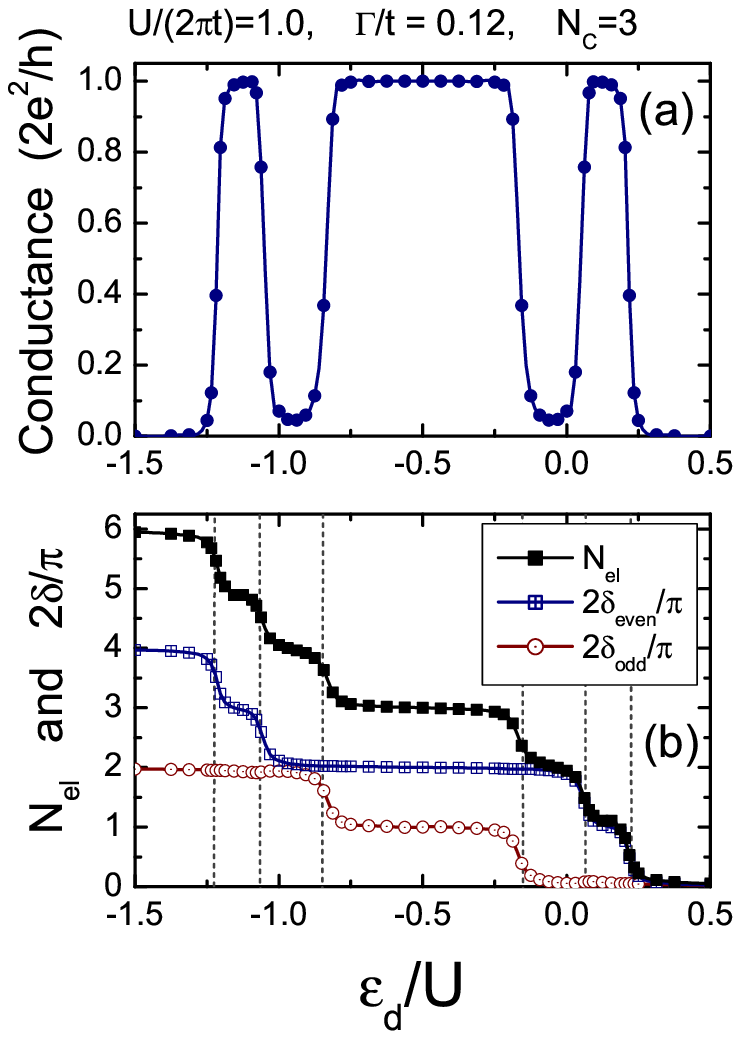}
\caption{
The conductance $g_{\rm series}^{\phantom{0}}$, local charge $N_{\rm el}$, 
phase shifts $2\delta_{\rm even}/\pi$ and  $2\delta_{\rm odd}/\pi$ 
 for the triple dot as a function of $\epsilon_d/U$ 
for $U/(2\pi t)=1.0$ and  $\Gamma/t = 0.12$. 
The dashed vertical lines in (b) show the values of $\epsilon_d$
at which $N_{\rm el}$ jumps in the \lq\lq molecule limit" $\Gamma = 0$.
For NRG, we use $t/D =0.1$ and $\Lambda=6.0$. 
}
\label{fig:cond_u1_g1}
\end{center}
\end{figure}

\begin{figure}[tb]
\begin{center}
\leavevmode
\includegraphics[ width=0.9\linewidth]{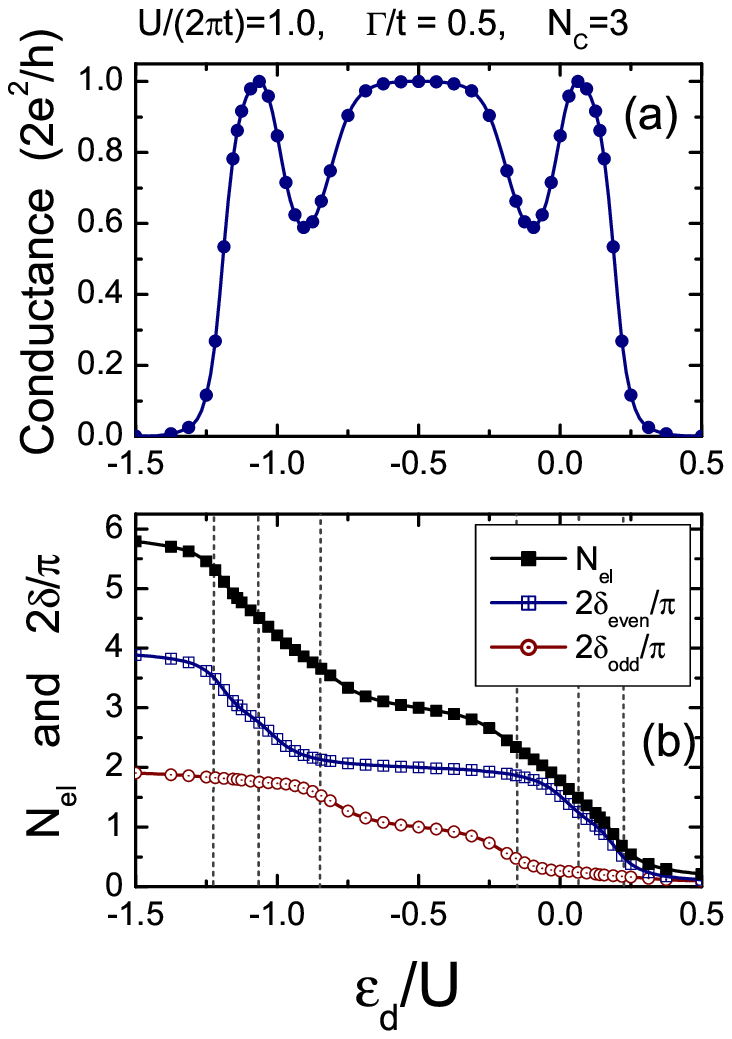}
\caption{
The conductance $g_{\rm series}^{\phantom{0}}$, local charge $N_{\rm el}$, 
phase shifts $2\delta_{\rm even}/\pi$ and  $2\delta_{\rm odd}/\pi$ 
for the triple dot as a function of $\epsilon_d/U$ 
for $U/(2\pi t)=1.0$ and  $\Gamma/t = 0.5$. 
The dashed vertical lines in (b) show the values of $\epsilon_d$
at which $N_{\rm el}$ jumps in the limit of $\Gamma = 0$.
For NRG, we use  $t/D =0.02$ and  $\Lambda=6.0$.    
}
\label{fig:cond_u1_g5}
\end{center}
\end{figure}

\begin{figure}[tb]
\begin{center}
\leavevmode
\includegraphics[ width=0.9\linewidth]{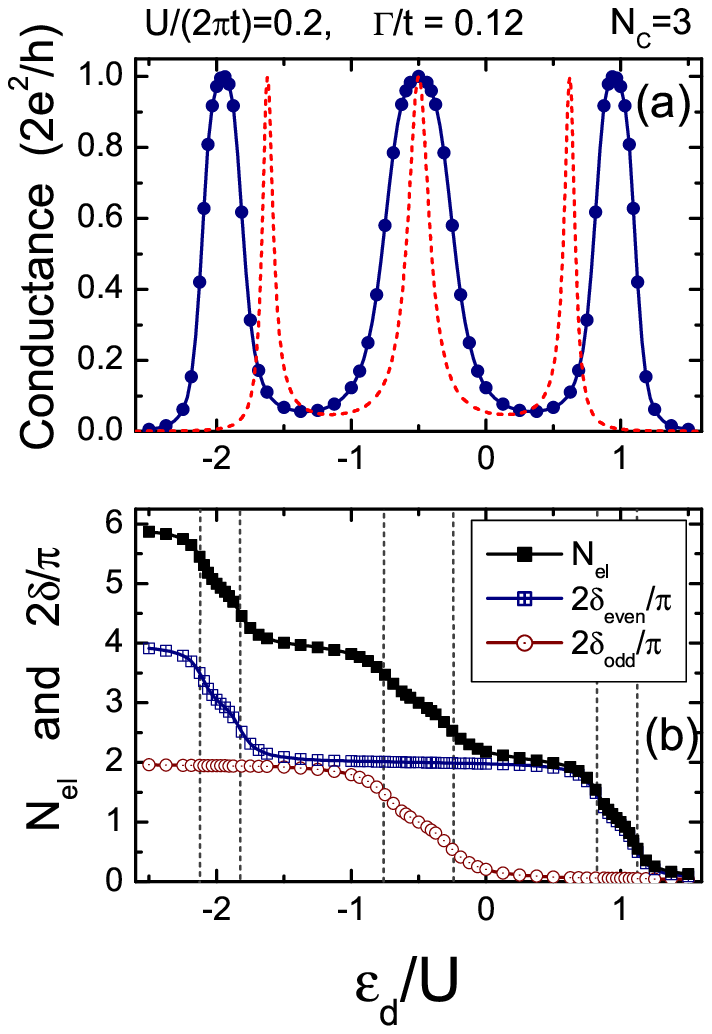}
\caption{
The conductance $g_{\rm series}^{\phantom{0}}$, local charge $N_{\rm el}$, 
phase shifts $2\delta_{\rm even}/\pi$ and  $2\delta_{\rm odd}/\pi$ 
for the triple dot as a function of $\epsilon_d/U$ 
for $U/(2\pi t)=0.2$ and  $\Gamma/t = 0.12$. 
In (a) the dashed line corresponds to the results for $U=0$:  
it is shifted to the negative energy region by $-U/2$ 
so that the position of the central peak coincides with that of 
the interacting case, 
and also the energy is  scaled by $U$ using $U/(2\pi t)=0.2$.
The dashed vertical lines in (b) show the values of $\epsilon_d$
at which $N_{\rm el}$ jumps in the limit of $\Gamma = 0$.
For NRG, we use  $t/D =0.1$ and $\Lambda=6.0$. 
}
\label{fig:cond_u02_g1}
\end{center}
\end{figure}

\subsection{Series connection}

We first of all discuss the noninteracting case $U=0$.
In Fig.\ \ref{fig:cond_u0},
the conductance $g_{\rm series}$,  
local charge in the triple dot $N_{\rm el}$,
and the phase shifts are 
plotted as functions of $\epsilon_d/t$ 
for (solid line) $\Gamma/t = 0.12$ and (dashed line) $\Gamma/t = 0.5$.
The conductance shows sharp peaks 
when the resonance states in the triple dot cross the Fermi level. 
Among the three conductance peaks, 
the one in the middle corresponds to an orbitals with the odd parity, 
and the remaining two belong to the orbitals with the even parity.
For $U=0$, the two phase shifts are given by  
$\cot \delta_{\rm even}  
= \left(\epsilon_d^2-2t^2\right)/\left( \Gamma \epsilon_d \right)$, 
and $\cot \delta_{\rm odd} = \epsilon_d/\Gamma$.
The phase shifts show a step of the height $\pi$,
when the resonance state of the corresponding parity 
crosses the Fermi level. Simultaneously, 
the local charge $N_{\rm el}$ also shows 
a staircase behavior, and the step corresponding 
to two electrons, spin up and spin down,
occupying the resonance states. 
For large $\Gamma$, the resonance peaks become broad, 
and the steps in the phase shifts vanish gradually.

For interacting case $U \neq 0$, 
we have carried out the NRG iterations 
by introducing the bonding and anti-bonding orbitals also for the leads;
$(f^{\dagger}_{n,R\sigma} \pm  f^{\dagger}_{n,L\sigma})/\sqrt{2}$. 
In the present study,
instead of adding these two orbitals at $n=N+1$ simultaneously 
for constructing the Hilbert space for the next NRG step,
we add the bonding orbitals first 
and retain typically 3600 low-energy states 
after carrying out the diagonalization. 
Then, we add the remaining anti-boding orbitals, 
and carry out the truncation again keeping the lowest 3600 eigenstates. 
This truncation procedure preserves the inversion symmetry, 
and in the case of odd $N_C$ it also preserves the particle-hole symmetry 
at $\epsilon_d = - U/2$. 
With this procedure,  
 the discretized Hamiltonian $H_N$ can be diagonalized exactly up to $N=1$, 
where the total number of the sites in the cluster 
is $N_C +2(N+1) = 7$ for the triple dot $N_C=3$ as 
illustrated in Fig.\ \ref{fig:model_NRG}. 
For $N \geq 2$, we have confirmed 
that 
the eigenvalues in the noninteracting limit $U \to 0$ 
are reproduced sufficiently well with this procedure for 
the discretization parameter of the value $\Lambda =6.0$.

We have also confirmed that the fixed-point eigenvalues of the NRG 
for the triple dot can be described by the quasi-particle Hamiltonian 
defined in eq.\ (\ref{eq:Hqp_even}) 
for all parameter values we have examined. 
It justifies the assumption of the local Fermi liquid 
we have made to derive eq.\ (\ref{eq:kappa2}).
In Fig.\ \ref{fig:cond_u1_g1}, 
the NRG results for (a) the conductance $g_{\rm series}^{\phantom{0}}$, 
(b) phase shifts $\delta_{\rm even}$, $\delta_{\rm odd}$ 
and total charge in the triple dot $N_{\rm el}$ 
are plotted as a function of $\epsilon_d/U$ 
for $U/(2\pi t) = 1.0$. Here the hybridization is relatively small
$\Gamma/t = 0.12$ compared to the hopping matrix element $t$,  
which is taken to be $t/D = 0.1$ in the NRG iterations.
The vertical dotted lines in (b)  
correspond to the values of $\epsilon_d$ 
at which $N_{\rm el}$ jumps discontinuously 
in a \lq molecule' limit $\Gamma \to 0$. 
The local charge $N_{\rm el}$ shows a step behavior 
near the dotted lines. Specifically, due to the Coulomb interaction, 
the steps emerge also for odd $N_{\rm el}$.
These odd steps reflect the $\pi/2$ steps of the phase shifts. 
The conductance shows the typical Kondo plateaus 
of the Unitary limit $g \simeq 2e^2/h$ in the regions 
of $\epsilon_d$ corresponding to 
the odd occupancies $N_{\rm el} \simeq 1$, $3$, $5$.
In contrast, the conductance shows wide minima   
for even occupancies $N_{\rm el}\simeq 2$, $4$. 
These features are linked to the behavior of the phase shifts; 
 $\delta_{\rm even} -\delta_{\rm odd} \simeq \pi/2$ 
for odd $N_{\rm el}$, and 
$\delta_{\rm even} -\delta_{\rm odd} \simeq 0$ or $\pi$ 
for even $N_{\rm el}$.\cite{Kawabata}

Among the three conductance plateaus, 
the one in the middle is wider than the others.
This seems to be caused by the suppression of  
the charge fluctuation near half-filling.
It is expected that 
the plateau becomes wider for some particular 
values of $N_{\rm el}$, at which the charging energy defined 
with respect to the unconnected limit $\Gamma=0$ is large. 
Thus, a wide plateau may emerge  
for the filling near the metal-insulator transition, 
which generally depends on the spatial range 
of the repulsive interaction.

We next consider the case where 
the triple dot is coupled strongly to 
the leads via a large hybridization.
In Fig.\ \ref{fig:cond_u1_g5},
the NRG results obtained at $\Gamma/t = 0.5$ are shown, 
where the value of the Coulomb interaction is unchanged $U/(2\pi t)=1.0$.
The local charge $N_{\rm el}$, 
which shows a staircase behavior for small $\Gamma$ 
in Fig.\ \ref{fig:cond_u1_g1}, 
becomes now a gentle slope in Fig.\ \ref{fig:cond_u1_g5} (b)
due to the large hybridization.  
Correspondingly, 
the shoulders of the conductance plateaus become round,
and the valleys become shallow.
Nevertheless, near half-filling, the phase shift 
for the odd partial wave still shows a weak $\pi/2$ step, 
i.e.,  $\delta_{\rm odd}\simeq \pi/2$ near half-filling.
It keeps the conductance peak at the center flat. 
Note that in this calculation 
the hopping matrix element is taken to be $t/D =0.02$, 
but the change in this ratio 
does not affects the low-energy properties so much.

We also examine the case with a weak interaction 
$U/(2\pi t)=0.2$ and small hybridization $\Gamma/t = 0.12$. 
The results are shown in Fig.\ \ref{fig:cond_u02_g1}.
It shows an early stage of a development of the Kondo plateaus.
In the upper panel (a),  
the dashed line corresponds to the conductance 
in the noninteracting case: for comparison 
it is shifted to the negative energy region by $-U/2$ 
so that the positions of the central peak coincides with 
that for the interacting case,
and also the energy is scaled by $U$ using the value $U/(2\pi t)=0.2$.
Due to the Coulomb interaction,  
the conductance peaks become wider than that for the noninteracting electrons,
and the form of the peaks deviates from the simple Lorentzian shape. 
Furthermore, the separation of the peaks increases. 
Since $U$ is not so large in this case,
the local charge $N_{\rm el}$ becomes flat only for   
even occupancies $N_{\rm el}=0,2,4,6$.
It reflects the behavior of the phase shifts, i.e.,
 both $\delta_{\rm even}$ and $\delta_{\rm odd}$ 
do not show the clear $\pi/2$ steps.
Nevertheless,  
the slopes in between the $\pi$ steps 
 become longer compared to the ones in the noninteracting case, 
and weak precursors for the development  
of the $\pi/2$ steps can be seen in these slopes.

\subsection{Parallel conductance}

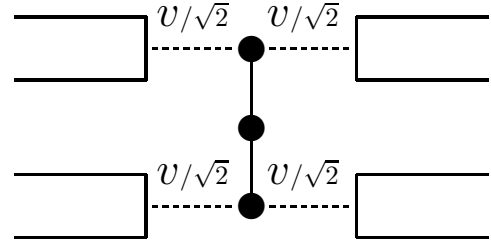
\begin{figure}[bt]
\begin{center}
\setlength{\unitlength}{0.7mm}

\begin{picture}(130,50)(-5,0)
\thicklines

\put(9,4){\line(1,0){25}}
\put(9,16){\line(1,0){25}}
\put(34,4){\line(0,1){12}}

\put(74,4){\line(1,0){25}}
\put(74,16){\line(1,0){25}}
\put(74,4){\line(0,1){12}}

\put(9,34){\line(1,0){25}}
\put(9,46){\line(1,0){25}}
\put(34,34){\line(0,1){12}}

\put(74,34){\line(1,0){25}}
\put(74,46){\line(1,0){25}}
\put(74,34){\line(0,1){12}}

\multiput(35.2,10)(2,0){8}{\line(1,0){1}}
\multiput(57.5,10)(2,0){8}{\line(1,0){1}}

\multiput(35.2,40)(2,0){8}{\line(1,0){1}}
\multiput(57.5,40)(2,0){8}{\line(1,0){1}}

\put(54,10){\circle*{5}} 
\put(54,25){\circle*{5}} 
\put(54,40){\circle*{5}} 

\put(54,12){\line(0,1){12}}
\put(54,27){\line(0,1){12}}


\put(36,13){\makebox(0,0)[bl]{{\Large $v$}$/\sqrt{2}$}}
\put(36,43){\makebox(0,0)[bl]{{\Large $v$}$/\sqrt{2}$}}
\put(57,13){\makebox(0,0)[bl]{{\Large $v$}$/\sqrt{2}$}}
\put(57,43){\makebox(0,0)[bl]{{\Large $v$}$/\sqrt{2}$}}


\end{picture}
\caption{Schematic picture of a parallel connection for $N_C=3$.}
\label{fig:parallel_3s}
\end{center}
\end{figure}

The parallel conductance through the tripled dot
as illustrated in Fig.\ \ref{fig:parallel_3s}, 
can be evaluated via eq.\ (\ref{eq:cond_para2}) with the phase shifts 
for the series connection given in the above.
This is because, as mentioned in \S \ref{sec:MODEL},
the matrix Green's function $\mbox{\boldmath $G$}(z)$ for 
the interacting sites for the parallel connection  coincides with 
that for the series connection 
owing to the inversion symmetry for the leads 
in the horizontal direction, as described in Fig.\ \ref{fig:parallel}.
Thus a site-diagonal quantity  
such as local charge $N_{\rm el}$ becomes  
the same with that for the series connection as mentioned 
in \S \ref{sec:MODEL}.
The conductances, however, are different.
It opens a way to determine the values 
the phase shifts $\delta_{\rm even}$ and $\delta_{\rm odd}$ 
from measurements of the series and parallel conductances  
using eqs.\  (\ref{eq:g_sym}) and (\ref{eq:cond_para2}).

\begin{figure}[tb]
\begin{center}
\leavevmode
\includegraphics[ width=0.9\linewidth]{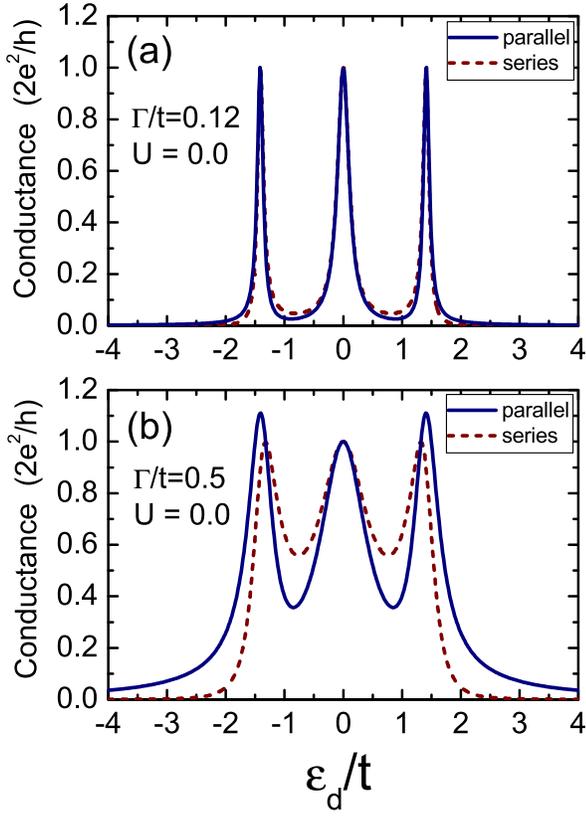}
\caption{
The conductance in the parallel (solid line) and series (dashed line) 
connections for noninteracting electrons $U=0$ as functions 
of $\epsilon_d/t$, where (a) $\Gamma/t=0.12$ and  (b) $\Gamma/t=0.5$. 
Note that two conducting channels contribute to  
the parallel conductance, so that the upper bound for
$g_{\rm parallel}$ is not $2e^2/h$.
}
\label{fig:cond_para_u0}
\end{center}
\end{figure}

We first of all consider the noninteracting case. 
In Fig.\ \ref{fig:cond_para_u0}, 
the parallel (solid line) and series (dashed line) 
conductances for $U=0$ are plotted as functions of $\epsilon/t$ 
for (a) $\Gamma/t=0.12$ and (b) $\Gamma/t=0.5$.  
In (a), the two curves almost overlap each other,
and the difference between the parallel and series 
conductances can be seen only near the two valleys 
corresponding to even $N_{\rm el}$.
The both conductances show a similar feature 
due to the three resonance states. 
At half-filling, $\epsilon_d = 0$ for noninteracting electrons,
the two curves contact with each other.
This is because the phase shifts take 
the values  $\delta_{\rm even}=\pi$ and  $\delta_{\rm odd}=\pi/2$. 
The phase shifts do not show such an exact synchronism 
at the side resonance peaks corresponding to $N_{\rm el} \simeq 1$ and $5$, 
and there the parallel conductance become larger than $2e^2/h$,
which is possible because two conducting channels 
contribute to the current in the parallel connection. 
The parallel conductance becomes larger 
than the series conductance for $|\epsilon_d| \gtrsim 2t$.
This is because the both phase shifts show 
the same asymptotic behavior, 
$\delta_{\rm even} \simeq \Gamma/\epsilon_d$ and 
$\delta_{\rm odd} \simeq \Gamma/\epsilon_d$, for $\epsilon_d \to \infty$,
and a similar cancellation occurs for the series conductance 
also in the opposite limit $\epsilon_d \to -\infty$.
The difference between the parallel and series conductances 
becomes small for small  $\Gamma$.

\begin{figure}[tb]
\begin{center}
\leavevmode
\includegraphics[ width=0.9\linewidth]{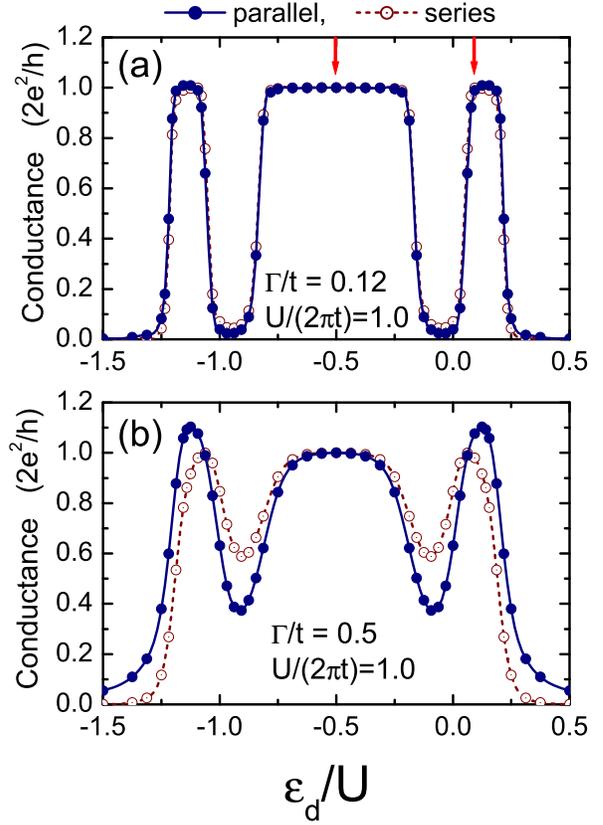}
\caption{
The conductance in the parallel (solid line) and series (dashed line) 
connections for interacting electrons $U/(2 \pi t)=1.0$ 
as functions of $\epsilon_d/U$, 
where (a) $\Gamma/t=0.12$ and  (b) $\Gamma/t=0.5$. 
The two arrows in (a) indicate the points, 
for which the energy flow is described in Fig.\ \ref{fig:flow}.
 }
 \label{fig:para_gam}
 \end{center}
 \end{figure}

In Fig.\ \ref{fig:para_gam},
 the (solid circle) parallel 
and (open square) series conductances 
in the interacting case $U/(2\pi t)=1.0$ 
are plotted as a function of $\epsilon_d/U$ 
for 
(a) $\Gamma/t = 0.12$ and 
(b) $\Gamma/t = 0.5$.
The difference between $g_{\rm parallel}$ and $g_{\rm series}$ 
for interacting electrons is similar, qualitatively, 
to that in the noninteracting case, 
although the appearance of the Kondo plateaus changes the 
overall feature of the $\epsilon_d$ dependence. 
In (a) the difference are visible only near the conductance valleys.
For large $\Gamma$, however, the difference becomes larger, 
and it can be seen whole region plotted in (b) except near 
the central plateau. As in the noninteracting case, the parallel 
conductance becomes larger than $2e^2/h$ at the side Kondo plateaus 
corresponding to $N_{\rm el} \simeq 1$ and $3$ due to the contributions 
of the two conducting modes.
The conductance valleys corresponding to even $N_{\rm el}$ become  
deeper for the parallel conductance than that of the series conductance.
However, the difference is small for small $\Gamma$.

\section{Discussion}
\label{sec:discussion}

The results presented in the above have been obtained at zero temperature. 
Nevertheless the Kondo behavior we have discussed can be seen 
at low enough temperatures $T \lesssim T_K$, although 
the height of the Kondo plateaus will decrease gradually 
with increasing $T$ from the Unitary-limit value as that 
in the single impurity case.\cite{IzumidaK40}
At $T \gg T_K$, the plateaus finally disappear and 
the conductance shows the oscillatory behavior 
of the Coulomb blockade with the  peaks,  
the height of which is about half of the Unitary-limit value.\cite{Kawabata}
The Kondo temperature $T_K$ decreases with increasing $U$,
and a large number of the NRG iteration $N$  
is required to get to the Fermi-liquid fixed point 
that determines the low-energy Kondo behavior for large $U$.\cite{ao_ah}
The value of $T_K$ should be different for different plateaus.
Specifically, for the central plateaus near half-filling, 
$T_K$ becomes small for the Hubbard chain with large size $N_C$ 
because the developing Mott-Hubbard (pseudo) gap disturbs 
 the screening of the local moment.

\begin{figure}[bt]
\begin{center}
\leavevmode
\includegraphics[ width=1.0\linewidth]{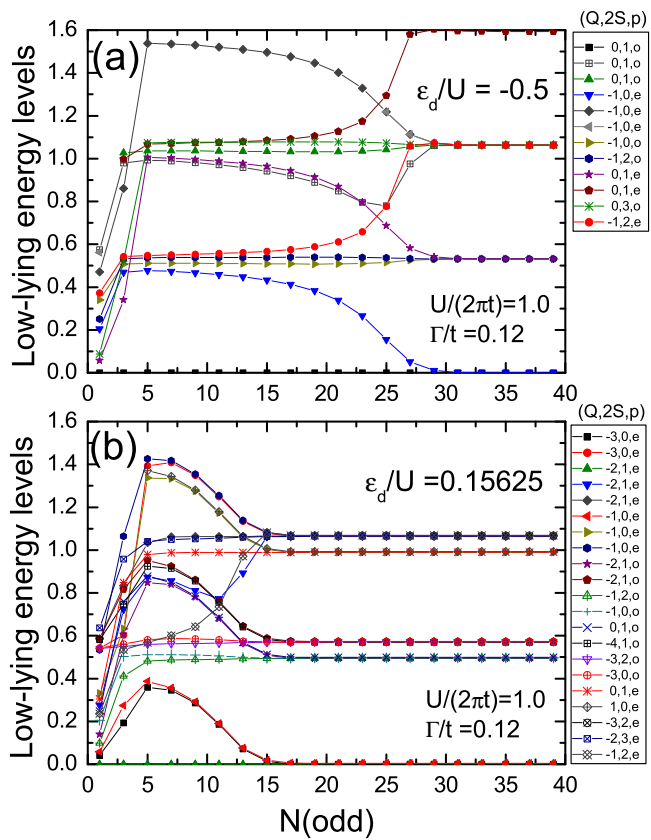}
\caption{
 Low energy levels of $H_N/D$ 
are plotted as a function of odd $N$ 
for (a) $\epsilon_d/U= -0.5$ where $N_{\rm el} =3$,
  and (b) $\epsilon_d/U= 0.15625$ where $N_{\rm el} \simeq 1$.
Other parameters are the same 
with those in the case of Fig.\ \ref{fig:cond_u1_g1}: 
 $U/(2\pi t)=1.0$   $\Gamma/t = 0.12$,
 $t/D =0.1$ and $\Lambda=6.0$. 
The label ($Q$, $2S$, $p$) corresponds to 
the charge $Q$, spin $S$, and parity $p$ ($=$ even, or odd).
}
\label{fig:flow}
\end{center}
\end{figure}

In order to estimate the Kondo energy scale for different plateaus,
the flow of the low-lying eigenvalues of $H_N/D$ 
for the triple-dot is plotted 
in Fig.\ \ref{fig:flow} as a function of odd $N$
for $U/(2 \pi t)= 1.0$ and $\Gamma/t=0.12$. 
This parameter set is the same as the one used for 
Fig.\ \ref{fig:cond_u1_g1} and Fig.\ \ref{fig:para_gam} (a).
The figure \ref{fig:flow} (a) shows 
the flow at the central Kondo plateau 
 for $\epsilon_d/U =-0.5$,   
and the lower panel (b) shows the flow at 
the right plateau for $\epsilon_d/U \simeq 0.15625$.
The positions corresponding to 
these two values of $\epsilon_d$  
are indicated in Fig.\ \ref{fig:para_gam} (a) by the arrows.
In the figure \ref{fig:flow}, 
the crossover from the high-energy region 
to low-energy Fermi-liquid regime can be seen clearly.
For large $N$, the energy levels converge 
to the fixed-point values, 
and the many-body eigenvalues of $H_N/D$ have one-to-one correspondence 
with the free quasi-particle excitations.\cite{ao_ah,KWW,KWW2}
The number of NRG iterations $N^*$ needed 
to reach the low-energy regime in each of the plateaus 
is estimated to be (a) $N^* \simeq 30$,  (b) $N^* \simeq 15$,
and  $N^* \simeq 15$ also for the left plateau,
for which the energy flow is identical to Fig.\ \ref{fig:flow} (b) except 
for the values of some quantum numbers assigned to the lines. 
The crossover occurs at an energy $T^* \simeq D\Lambda^{-(N^*-1)/2}$,
where a factor of the form $\Lambda^{-(N-1)/2}$ emerges 
to recover the energy scale of the original Hamiltonian ${\cal H}$ from 
the discretized version $H_N$ defined in eq.\ (\ref{eq:H_N}). 
The crossover energy reflects the width of the Kondo resonance, 
so that  $T^* \sim T_K$ apart form a numerical factor 
of order O(1) which depends on the precise definition of $T_K$. 
Consequently, these results show that 
the Kondo energy scale for the central plateau 
at half-filling is much smaller than 
that for the side plateaus away from half-filling.

In order to calculate the conductance for whole temperature ranges,
one generally has to use several different approaches.
The behavior at high temperatures $T > T_K$ can be captured,  
for instance, by the exact diagonalization of small clusters, 
such as the one examined by several authors.\cite{Chiape,Busser}
At low temperatures, however, 
one needs the information about the low-energy excitations. 
From the Fermi-liquid regime to the crossover region near $T_K$, 
the conductance can be calculated from 
the current-current correlation function 
using NRG method.\cite{Izumida2,IzumidaK40}
The ground-state value of the conductance 
can be calculated from the phase shifts, 
which can be deduced efficiently 
from the low-energy eigenvalues using eq.\ (\ref{eq:kappa2}). 

In summary, we have described the method to 
deduce the phase shifts for finite 
interacting systems attached to noninteracting leads 
from the fixed point Hamiltonian of NRG.
Our approach assumes the inversion symmetry,
 but does not assume a specific form of the Hamiltonian 
nor electron-hole symmetry, and thus
it is applicable to a wide class of quantum impurities. 
We apply the method to a triple quantum dot of $N_C=3$ connected 
to two noninteracting leads, 
and calculate the dc conductance away from half-filling 
as a function of the onsite energy $\epsilon_d$.
At $T=0$, the conductance shows the Kondo plateaus 
of the Unitary limit $g \simeq 2e^2/h$ at the values of gate 
voltage corresponding to odd numbers of electrons $N_{\rm el}$, 
while it shows the wide minima for even $N_{\rm el}$. 
It seems to be natural to expect that 
these low-temperature properties seen at $T \lesssim T_K$ 
are common to the Hubbard chain of finite size $N_C$ attached 
to Fermi-liquid reservoirs.

\bigskip

\section*{Acknowledgements}
One of us (ACH) wishes to thank the EPSRC(Grant GR/S18571/01) 
for financial support.
Numerical computation was partly performed 
at Yukawa Institute Computer Facility.


\begin{thebibliography}{99}



\bibitem{GlatzmanRaikh}
L. I. Glazman and M. E. Raikh: 
JETP Lett.\ {\bf 47} (1988) 452.


\bibitem{NgLee}
T. K. Ng and P. A. Lee: Phys.\ Rev.\ Lett.\ {\bf 61} (1988) 1768.


\bibitem{Goldharber}
D. Goldharber-Gordon, H. Shtrikman, D. Mahalu, D. Abusch-Magder,
U. Meirav, and M. A. Kastner:
Nature {\bf 391} (1998)  156.


\bibitem{Cronenwett}
S. M. Cronenwett, T. H. Oosterkamp, and L. P. Kouwenhoven: 
Science {\bf 281} (1998) 540.


\bibitem{Hofstetter}
W. Hofstetter, J. K\"{o}nig, and H. Schoeller:
Phys. Rev. Lett. {\bf 87} (2001) 156803.


\bibitem{Kobayashi}
K. Kobayashi, H. Aikawa, S. Katsumoto, and Y. Iye:
Phys. Rev. Lett. {\bf 88}  (2002) 256806.



\bibitem{SC}
A. Oguri, Y. Tanaka, and A. C. Hewson:
J.\ Phys.\ Soc.\ Jpn.\   {\bf 73} (2004) 2494. 



\bibitem{Wilson}
K. G. Wilson: Rev.\ Mod.\ Phys.\  {\bf 47} (1975) 773. 


\bibitem{KWW}
H. R. Krishna-murthy, J. W. Wilkins: 
and K. G. Wilson, Phys.\ Rev.\ B {\bf 21} (1980) 1003. 

\bibitem{KWW2}
H. R. Krishna-murthy, J. W. Wilkins: 
and K. G. Wilson, Phys.\ Rev.\ B {\bf 21} (1980) 1044. 




\bibitem{Izumida2}
W. Izumida, O. Sakai, and Y. Shimizu: 
 J.\ Phys.\ Soc.\ Jpn.\ {\bf 67} (1998) 2444.




\bibitem{IzumidaK40}
W. Izumida and O. Sakai: 
 Phys.\ Rev.\ B {\bf 62} (2000) 10260;
 J.\ Phys.\ Soc.\ Jpn.\  {\bf 74} (2005) 103.


\bibitem{ao_ah}
A. Oguri and A. C. Hewson: 
J.\ Phys.\ Soc.\ Jpn.\ {\bf 74} (2005) 988.




\bibitem{HewsonOguriMeyer}
A. C. Hewson, A. Oguri, and D. Meyer: 
Eur.\ Phys.\ J.\ B {\bf 40} (2004) 177.

\bibitem{Hewson_JPSJ}
A. C. Hewson: J.\ Phys.\ Soc.\ Jpn.\  {\bf 74} (2005) 8.  



\bibitem{Kawabata}
A. Kawabata, J.\ Phys.\ Soc.\ Jpn.\ {\bf 60}, 3222 (1991).


\bibitem{ao8}
A. Oguri:  Phys.\ Rev.\ B {\bf 59}  (1999) 12240. 

\bibitem{ao9}
A. Oguri: Phys. Rev. B {\bf 63}  (2001) 115305; 
{\em ibid.\/} [Errata: {\bf 63} (2001) 249901].  


\bibitem{ao10}
A. Oguri: 
J.\ Phys.\ Soc.\ Jpn.\ \textbf{70} (2001) 2666;
{\em ibid.\/} {\bf 72},  3301 (2003).




\bibitem{LangerAmbegaokar}
J. S. Langer and V. Ambegaokar, 
Phys.\ Rev.\ {\bf 121} (1961) 1090.



\bibitem{tanaka1}
Y. Tanaka, A. Oguri, and H. Ishii:
J.\ Phys.\ Soc.\ Jpn.\ {\bf 71}  (2002) 211.




\bibitem{Hewson_jphys}
A. C. Hewson: J.\ Phys.:\ Condens.\ Matter {\bf 13} (2001) 10011.









\bibitem{SakaiShimizuKasuya}
O. Sakai, Y. Shimizu, and T. Kasuya:
Prog.\ Theor.\ Phys.\ Suppl.\ {\bf 108} (1992) 73.



\bibitem{Chiape}
G. Chiappe and J. A. Verg\'{e}s: 
J.\ Phys.:\ Condens.\ Matter {\bf 15} (2003) 8805.


\bibitem{Busser}
C. A. B\"{u}sser, A. Moreo, and E. Dagotto: 
Phys. Rev. B 70  (2004)  035402. 








\end{thebibliography}
\end{document}